\documentclass[12pt]{article}
\usepackage{osajnl}

\input{tcilatex}

\begin{document}

\title{Exact nonparaxial beams of the scalar Helmholtz equation}
\author{G. Rodr\'{\i}guez-Morales, S. Ch\'{a}vez-Cerda \\
\address{ Instituto Nacinal de Astrof\'{\i}sica, Optica y Electr\'{o}nica
 Apdo Postal 51/216, Puebla, Pue. 72000, M\'{e}xico}\bigskip \bigskip }

\begin{abstract}
It is shown that three-dimensional nonparaxial beams are described by the
oblate spheroidal exact solutions of the Helmholtz equation. For the first
time, their beam behaviour is investigated and their corresponding
parameters are defined. Using the fact that the beam width of the family of
paraxial Gaussian beams is described by an hyperbola, the connection between
the physical parameters of nonparaxial spheroidal beam solutions and those
of paraxial beams is formally stablished. These results are also helpful to
investigate the exact vector nonparaxial beams.
\end{abstract}

\maketitle


\newpage

Recent developments in science and technology brings the necessity of
revisiting the theory and concepts of nonparaxiality of optical beams. Its
understanding has great relevance in the description of optical fields which
are tightly focused or, optical beams whose diameter can be of the order of
a few wavelengths as can be the case in optical nanolasers.

When studying beams beyond the paraxial approximation, the standard
approach, introduced by Lax, has been to start with the paraxial solution
and then to include some corrections \cite{lax,seshadri,ciattoni,cao}.
Rigurously speaking, nonparaxial beams are solutions of the wave equation
without the paraxial approximation, in other words, they are solutions of
the Helmholtz wave equation $\nabla ^{2}E+k^{2}E=0$, with $k$ the wave
number. Thus, such kind of solutions must be investigated in order to
describe nonparaxial beams.

A class of nonparaxial solutions has been introduced assuming a point source
located at a complex position along the $z-$axis. The \emph{spherical} wave
field from this virtual complex source is converted into a directional wave
field which remains a rigurous solution of the Helmholtz equation \cite{shin}%
. This virtual source mode has wavefronts that are approximately oblate
spheroidal \cite{shep}. This fact has also been discussed by several authors
and even have expressed the virtual source mode wave in spheroidal
coordinates \cite{landbar,ulan,saari,kraus}. Whatever the coordinate system
used, the field expression reduces, under the paraxial approximation, to
Gaussian beam \cite{shep,landbar,ulan,saari,kraus,deschamps}.

The virtual complex source point solutions carry an inherent singularity
which makes them inadequate to describe propagating fields near the origin
or focal source point \cite{shep,ulan}. To elliminate this problem a
non-singular superposition of incoming and outgoing spherical waves has been
used \cite{shep,ulan}, however even this non-singular solution has problems
since, to realise them physically, infinite energy is required \cite{lekner}.

In this letter we investigate the solutions of the scalar Helmholtz equation
in oblate spheroidal coordinates and we show that they can have a beam
behaviour. By identifying that the evolution of the beam width of a Gaussian
mode follows an hyperbola we relate the physical parameters of nonparaxial
spheroidal beam solutions and those of paraxial beams. This also allows to
define quantitatively a threshold between nonparaxial and paraxial beams. In
the paraxial limit the oblate spheroidal solution tend to Laguerre-Gaussian
beams.

Pure plane, spherical or cylindrical waves of the Helmholz equation, cannot
be used, in a strict sense, to describe optical beams if we understand as
such those field distributions that are concentrated around an imaginary
line that will be identified as the beam propagation axis. Here we show that
solutions of the Helmholtz equation in spheroidal coordinates can satisfy
this requirement depending on the value of the ellipsoidal parameter (to be
defined below).

The propagation axis will be the $z-$axis that will also be the azimuthal
axis of the oblate spheroid coordinate system. This coordinates are defined
by $x=d(\xi ^{2}+1)^{1/2}(1-\eta ^{2})^{1/2}\cos \phi $, $y=d(\xi
^{2}+1)^{1/2}(1-\eta ^{2})^{1/2}\sin \phi $ and $z=d\xi \eta $, where $d>0$
is the distance from the origin to the foci \cite{flammer,wei}. The
variation range of the coordinates is determined by the physical problem. In
the present case they are $0\leq \eta \leq 1$, $-\infty \leq \xi \leq \infty 
$ and $0\leq \phi \leq 2\pi $. The first two coordinates can be related to
an eccentric radial coordinate $\rho =\sinh ^{-1}\xi $ and an eccentric
angular coordinate by $\theta =\cos ^{-1}\eta $. From the above coordinate
transformations, the condition $|\eta |=1$ defines the $z-$axis.

In oblate spheroidal coordinates $(\xi ,\eta ,\phi )$ the Helmholtz wave
equation $\nabla ^{2}E+k^{2}E=0$, is separated into the next set of
equations \cite{flammer,wei} 
\begin{equation}
\begin{array}{c}
\frac{d^{2}\Phi }{d\phi ^{2}}+m^{2}\Phi =0 \\ 
\left( \xi ^{2}+1\right) \frac{d^{2}R}{d\xi ^{2}}+2\xi \frac{dR}{d\xi }%
-\left( a_{mn}-c^{2}\xi ^{2}+\frac{m^{2}}{\xi ^{2}+1}\right) R=0 \\ 
\left( 1-\eta ^{2}\right) \frac{d^{2}S}{d\eta ^{2}}-2\eta \frac{dS}{d\eta }%
+\left( a_{mn}-c^{2}\eta ^{2}-\frac{m^{2}}{1-\eta ^{2}}\right) S=0
\end{array}
\label{OblateHelmholtz}
\end{equation}
where $c=kd$ with $a_{mn}$ is a constant parameter. The spheroidal parameter 
$c$ is a meassure of how far is the spheroid from a sphere for which $d=0$
and in consequence $c=0$. From Eq. (\ref{OblateHelmholtz}) the solutions of
Helmholtz equation have the form of a Lam\'{e} product \cite{flammer} 
\begin{equation}
E_{nm}\left( \xi ,\eta ,\phi \right) =R_{mn}^{\left( p\right) }\left( \xi
;c\right) S_{mn}\left( \eta ;c\right) e^{im\phi }  \label{solsoblate}
\end{equation}
where in the spheroidal radial functions $R_{mn}^{(p)}\left( \xi ;c\right) $%
, the superindex $p$ defines the kind of solution (first kind $p=1$, second
kind $p=2$ and so on). With $p=3,4$ the solutions are linear combinations of
the first two, namely $R_{mn}^{\left( 3,4\right) }=R_{mn}^{\left( 1\right)
}\pm iR_{mn}^{\left( 2\right) }$, and these are what we will use as they,
can be used to describe travelling waves. There are two spheroidal angular
solutions $S_{mn}^{\left( 1,2\right) }\left( \eta ;c\right) $ but, we
exclude the second one since it tends to infinity as it approaches the $z$%
-axis then, only that of first kind can describe beams of finite energy and
so we drop the superindex for this solution.

In order to understand the physics of Eq. (\ref{solsoblate}) we refer to
Fig. 1 where the intensity of the modes with $m=n=0$ and $m=n=1$ are shown.
One immediately realises that for $c\neq 0$, the wavefield distribution is
concentrated around the $z-$axis, in other words, it behaves like a \emph{%
beam}. The angular part of (\ref{solsoblate})\ determines the transverse
pattern of the wavefield while the radial part defines the propagation
features. The azimuthal and radial parts of the solution determine the form
of the propagating phase front, when $m=0$ they are oblate spheroidal, Fig
1a), and when $m\neq 0$ the solution describes wavefields with rotating
wavefronts, Fig 1b).

To relate spheroidal solutions, Eq. (\ref{solsoblate}) with those of the
paraxial wave equation analyse the geometry of paraxial beams. Their beam
width $w(z)$ is given by $w^{2}(z)=w_{0}^{2}(1+z^{2}/R_{0}^{2})$ where $w_{0}
$ is the beam waist and $R_{0}$ is the Rayleigh distance \cite{siegman}.
Rewriting this expression we get the hyperbola equation $%
w^{2}(z)/w_{0}^{2}-z^{2}/R_{0}^{2}=1$ to which we can associate an oblate
spheroidal coordinate system, this is shown in Fig. 2. In it, $d$ is the
distance to one of the foci and $w_{0}$ is the vertex of the hyperbola. From
the geometry of the hyperbola, the equation 
\begin{equation}
R_{0}^{2}=d^{2}-w_{0}^{2}  \label{pythagoras}
\end{equation}
has to be fulfilled. From the equation of the hyperbola, the angle of the
asymptote or far field angle is $\tan \alpha =w_{0}/R_{0}$, this assigns a
geometric place to $R_{0}$ in Fig. 2. We now make the connection between
paraxial and spheroidal beams through the common parameter $d$. Sustituting $%
c=kd$ and $R_{0}=kw_{0}^{2}/2$ in Eq. (\ref{pythagoras}), after some
algebra, we get 
\begin{equation}
w_{0}=\frac{\sqrt{2(-1+\sqrt{1+c^{2}})}}{k}.  \label{wandc}
\end{equation}
This equation is the link that connects oblate spheroidal beams and paraxial
beams. There must exist a range of values of $c$ for which oblate spheroidal
beams describe paraxial beams. This can be obtained, expressing this
parameter in terms of the far field angle $c=2/(\tan \alpha \sin \alpha )$,
since paraxial beams cannot diverge at cone angles larger than $\alpha
\approx 0.5$rad $\approx $ $\pi /6$ \cite{siegman} we have $c_{p}\approx 7$
this value defines the frontier between paraxial and nonparaxial beams. In
Fig. 3 we show the plots of the on-axis intensity $\left| R_{mn}^{\left(
3,4\right) }(\xi ;c_{p})\right| ^{2}$ and transverse profile $S_{mn}\left(
\eta ;c_{p}\right) $ of an oblate spheroidal beam in this limit and compare
them with the corresponding paraxial beam. Also, for reference, it is
plotted the normalised axial intensity for a highly nonparaxial beam with $%
c<c_{p}$ (inner thin line). The small differences can be attributed to that
the paraxial Laguerre-Gauss beam is in the limit of its validity. In fact,
for $c>c_{p}$ the behaviour of the oblate spheroidal beams is even closer to
paraxial beams. We have evaluated the wavefield profiles for large values of 
$c>c_{p}$ and found that the transverse part of the spheroidal beams
described by the angular solutions $S_{mn}\left( \eta ;c\right) $ tend to
the Laguerre-Gauss beam, as expected \cite{flammer}. This similarity occurs
earlier for the zero-order mode for $c$ just above $c_{p}$. Finally, notice
that from Eq. (\ref{wandc}) when $c\rightarrow \infty $ the radial functions 
$R_{mn}^{\left( 3,4\right) }$ tend to plane waves for the fundamental order
mode and when $c\rightarrow 0$, the nonparaxial oblate spheroidal beams tend
to resemble spherical waves.

That the solutions discussed here are different from those obtained with the
complex source point method (CSPM) expressed in spheroidal coordinates \cite
{landbar,ulan,saari,kraus}. The zero-order nonparaxial beams obtained with
the CSPM have a singularity inherited from its construction from a source
point \cite{lekner}. The oblate spheroidal beams presented here do not
present any singularity whenever $c\neq 0$. Using the CSPM\cite
{landbar,ulan,saari,kraus}, it is usually proposed that in the paraxial
limit $d$ can take the value of $R_{0}$. From eq. (\ref{pythagoras}) it is
clear that this situation can never be possible since it would imply a
physical inconsistency, $w_{0}=0$ and a far-field angle $\alpha =0$ (see
Fig. 2).

In conlcusion, for the first time, we have demonstrated that spheroidal
beams describe paraxial and nonparaxial beam exact solutions of the
Helmholtz equation. We have established a threshold that separates
nonparaxial and paraxial regimes. Higher oblate spheroidal modes reduce in a
natural way to Laguerre-Gauss beams when the spheroidal parameter $c$
exceeds the critical value $c_{p}$. The results presented here were in the
neighborhood of the paraxial limit and thus the scalar treatment is valid.
Our approach extends straightforward to vector nonparaxial beams in Eq. (\ref
{solsoblate}), to obtain the corresponding vectorial components obtained 
\cite{stratton}. Thorough investigations are underway.

Authors acknowledge to CONACyT. GRM Acknowledge support to CONACyT grant
118631/120329.


\newpage

\begin{center}
{\Large Figure Captions}
\end{center}

\begin{enumerate}
\item  \label{fig:coords} Intesity and phase evolution (left), and
transverse pattern (right) of oblate spheroidal modes for $c=7$; a) $m=n=0$
b) $m=n=1$.

\item  Hyperbolic-elliptic geometry of a paraxial beam defined by the beam
width evolution.

\item  a) Comparison of on-axis intensity for oblate spheroidal nonparaxial
(thick solid line) and paraxial (dashed line) beams. Also it is shown the
normalised intensity for a highly nonparaxial beam (thin line). b)
Corresponding transverse amplitude profiles at the plane $z=0$. Notice that
for the Laguerre paraxial mode $r$ can be obtained from $%
r^{2}=x^{2}+y^{2}=d^{2}\left( 1-\eta ^{2}\right) $.\newpage 
\end{enumerate}

\FRAME{ftbpF}{6.0407in}{4.5377in}{0pt}{}{}{fig1.jpg}{\special{language
"Scientific Word";type "GRAPHIC";maintain-aspect-ratio TRUE;display
"USEDEF";valid_file "F";width 6.0407in;height 4.5377in;depth
0pt;original-width 12.5in;original-height 9.3754in;cropleft "0";croptop
"1";cropright "1";cropbottom "0";filename 'fig1.jpg';file-properties
"XNPEU";}}

\FRAME{ftbpF}{5.5832in}{4.5359in}{0pt}{}{}{fig2.jpg}{\special{language
"Scientific Word";type "GRAPHIC";maintain-aspect-ratio TRUE;display
"USEDEF";valid_file "F";width 5.5832in;height 4.5359in;depth
0pt;original-width 11.5003in;original-height 9.3331in;cropleft "0";croptop
"1";cropright "1";cropbottom "0";filename 'fig2.jpg';file-properties
"XNPEU";}}

\FRAME{ftbpF}{6.0407in}{4.5377in}{0pt}{}{}{fig3.jpg}{\special{language
"Scientific Word";type "GRAPHIC";maintain-aspect-ratio TRUE;display
"USEDEF";valid_file "F";width 6.0407in;height 4.5377in;depth
0pt;original-width 12.5in;original-height 9.3754in;cropleft "0";croptop
"1";cropright "1";cropbottom "0";filename 'fig3.jpg';file-properties
"XNPEU";}}

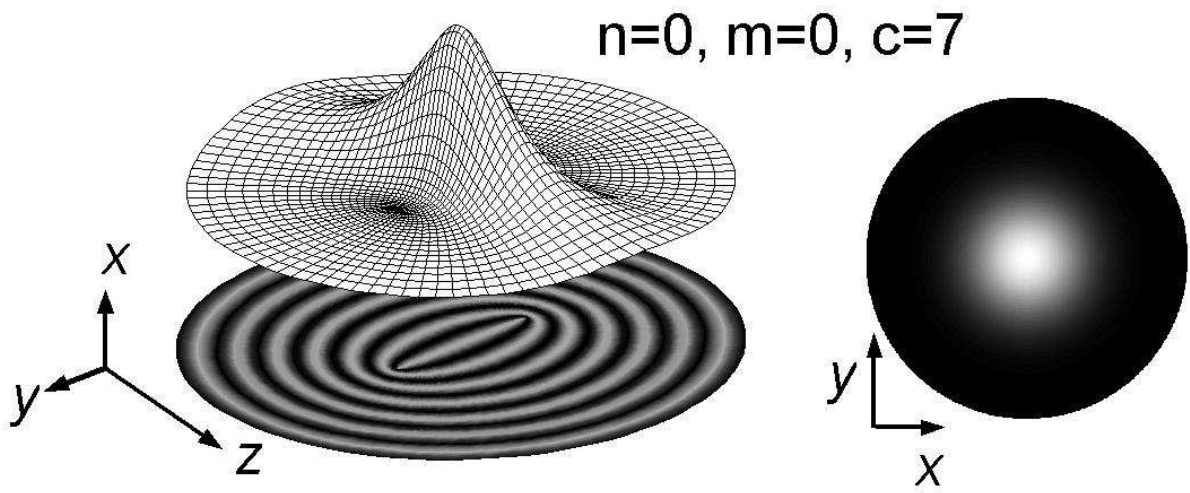
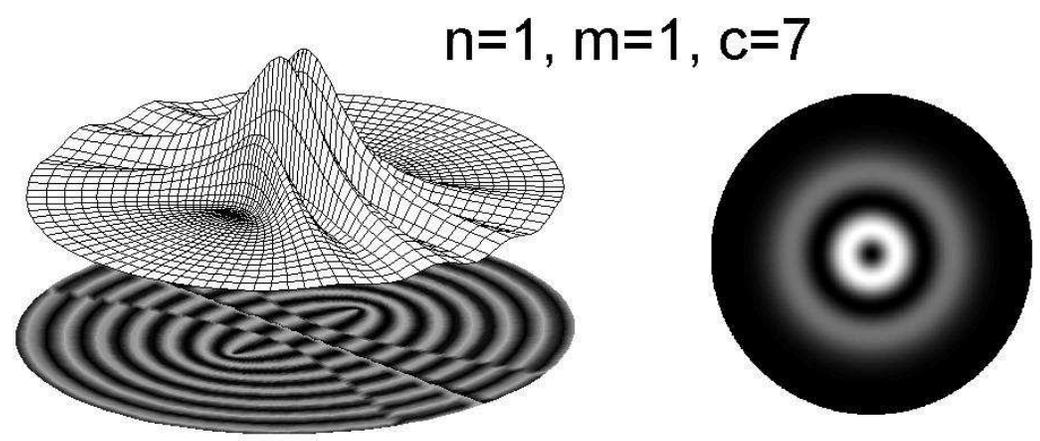
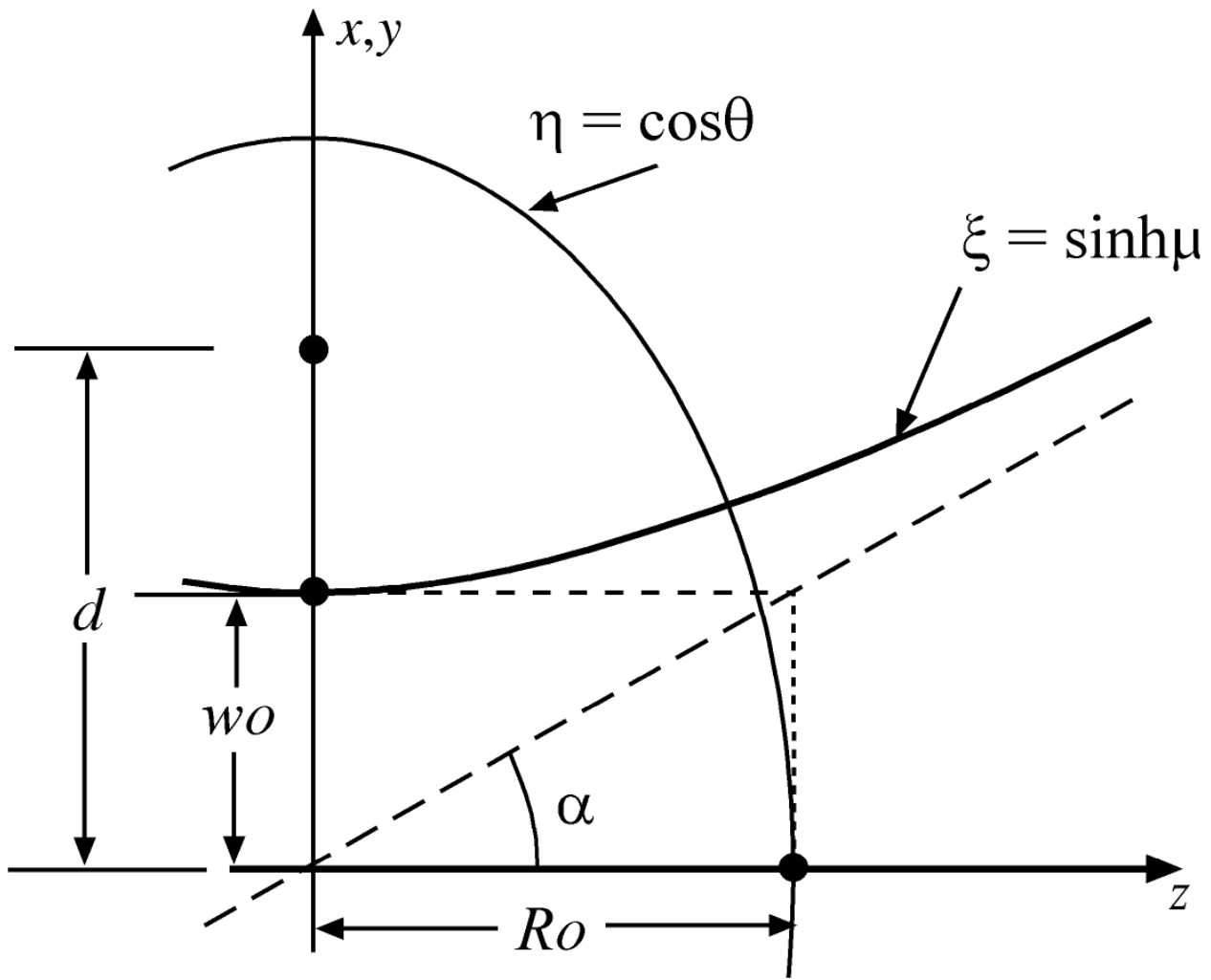
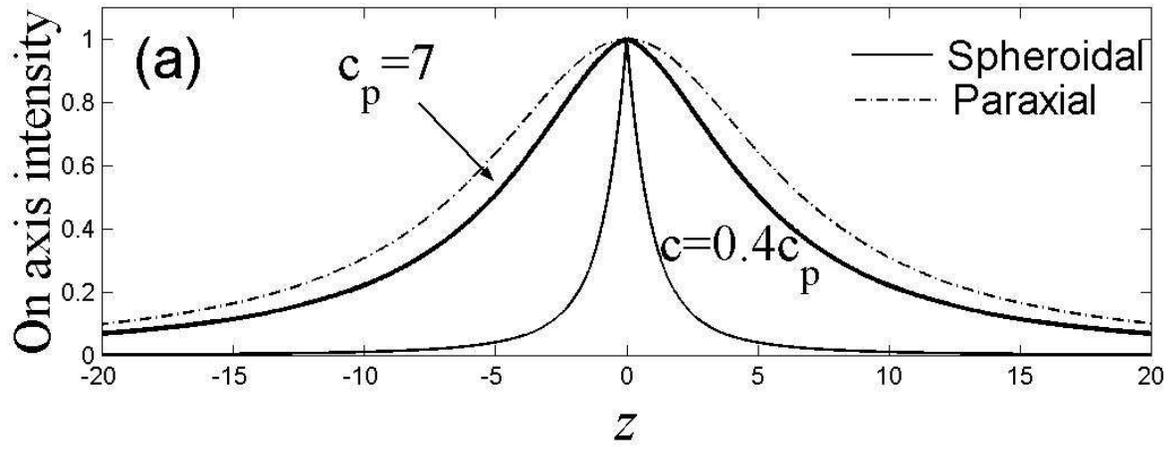
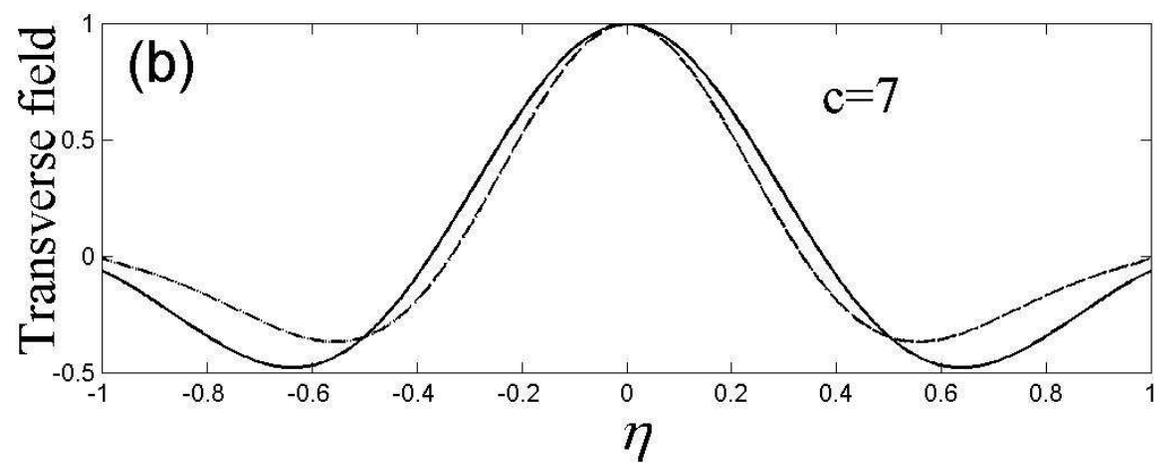
\end{document}